\documentclass[a4paper]{article}
\usepackage[british]{babel}
\usepackage{graphicx}

\begin{document}

\begin{center}
  {\Large \textbf{Separability and the stella octangula}}\\[14pt]
  \AA sa Ericsson\footnote{e-mail: x00aer@physto.se} \\[8pt]
  {\small Stockholm University, SCFAB\\
          Fysikum\\
      S-106 91 Stockholm, Sweden}\\[20pt]
  {\large \textbf{Abstract}} \\
\end{center}
{\small A geometrical picture of separability of ${2\times 2}$
        composite quantum systems, showing the region of separable
        density matrices in the space of hermitian matrices, is given.
        It rests on the criterion of separability given by Peres~[1],
    and it is an extension of the ``Horodecki diagram''~[2] and the
    ``stella octangula'' described by Aravind~[3].}

\vspace{10mm}

An illustration of which composite quantum states that are
separable is desirable. The separable states is a convex set in
the space of hermitian matrices, and a nice geometrical picture of
this for Bell-diagonal $2\times2$ systems is given by the
``Horodecki diagram''~[2]. Here we will extend this idea to
non-Bell-diagonal mixtures.

A composite quantum system of two subsystems, described by a
density matrix $\rho$ in the Hilbert-Schmidt space
$\mathcal{H}\otimes\mathcal{H}$, is separable if it can be written
as a statistical mixture of product states,
\begin{equation}
  \rho=\sum_k w_k \left( \rho_a \right)_k \otimes \left( \rho_b \right)_k \ ,
\end{equation}
where $w_k$ are positive with $\sum_k w_k=1$, and where $\rho_a$
and $\rho_b$ are density matrices for the two subsystems. A
necessary criterion for separability, given by Peres~[1], is that
the partial transpose $\rho^{T_b}$ of the density matrix is again
a density matrix, i.e. has only non-negative eigenvalues:
\begin{equation}
  \rho^{T_b}_{m\mu,n\nu}= \rho_{m\nu,n\mu} \ ,\qquad \rho^{T_b} \geq 0
  \ .
\end{equation}
This criterion has been shown also to be sufficient for
separability of $2\times2$ and $2\times3$ systems~[4].

In the Horodecki diagram the Bell basis states,
$\Psi^{\pm}=|\!\uparrow\downarrow\rangle \pm
|\!\downarrow\uparrow\rangle$ and
$\Phi^{\pm}=|\!\uparrow\uparrow\rangle \pm
|\!\downarrow\downarrow\rangle$, are at the vertices of a regular
tetrahedron, in which all other states of a $2\times2$ system lie
[1]. (The cartesian coordinates for a density matrix $\rho$ in the
Horodecki diagram is given by the eigenvalues of the correlation
tensor $T_{nm}=\mathrm{Tr}(\rho\:\sigma_n\otimes\sigma_m)\,.\,$)
In the tetrahedron there is an octahedron containing every
Bell-diagonal mixture that is separable. Aravind has pointed out
that in the Horodecki diagram, Peres' transformation---taking the
partial transpose---amounts to a reflection~[4]. The reflection of
the tetrahedron gives its inverse and the two tetrahedra form a
``stella octangula''. The region they have in common is the
octahedron. Thus the Bell-diagonal states in the octahedron are
separable because their partial transpose stays in the
tetrahedron. The states outside the octahedron will go to
unphysical regions outside the tetrahedron and are therefore
non-separable. But non-Bell-diagonal mixtures and their partial
transposes depend on additional parameters and can not be
satisfactory illustrated in the Horodecki diagram. To get a
picture of some of these states we will generalize to similar
tetrahedra formed by other basis states than the Bell basis. In
these tetrahedra the region of separable states will be found.

The density matrices of $2\times 2$ systems form a convex set in the
15 real dimensional flat space of all hermitian $4\times 4$
matrices with unit trace. With the Hilbert-Schmidt distance
$D=\sqrt{{\scriptstyle \frac{1}{2}} \mathrm{Tr}(\rho_1-\rho_2)^2}$
we get unit distance between pure orthogonal states. This means
that the density matrices $\rho_i \, ,\ i=1, 2, 3, 4$,  of four
basis states $|i\rangle$ will sit at the vertices of a regular
tetrahedron which spans a 3-dimensional subspace. The mixtures
$\rho=\sum_i w_i\rho_i \, ,\ \sum_i w_i=1 \,$, lie within the
tetrahedron. As basis states we choose:
\begin{equation}
  \left. \begin{array}{rcl}
  |1\rangle \!\!\!&=&\!\!\! \cos \alpha |\!\uparrow\uparrow\rangle + \sin \alpha |\!\downarrow\downarrow\rangle\\
  |2\rangle \!\!\!&=&\!\!\! \sin \alpha |\!\uparrow\uparrow\rangle - \cos \alpha |\!\downarrow\downarrow\rangle\\
  |3\rangle \!\!\!&=&\!\!\! \cos \alpha |\!\uparrow\downarrow\rangle + \sin \alpha |\!\downarrow\uparrow\rangle\\
  |4\rangle \!\!\!&=&\!\!\! \sin \alpha |\!\uparrow\downarrow\rangle - \cos \alpha |\!\downarrow\uparrow\rangle
  \end{array} \right.
\end{equation}
with $0\leq \alpha \leq \pi/4$. These basis states are, for each
$\alpha$, equally entangled (i.e. they have the same von Neumann
entropy $-\textrm{Tr}(\rho_a\textrm{ln}\rho_a)$, for the partial
traced matrices $\rho_a=\textrm{Tr}_b\rho$). For $\alpha=0$ we
have the separable product basis states. The entanglement
increases with $\alpha$ and for $\alpha=\pi/4$ we have the
maximally entangled Bell basis. $\alpha=\pi/4$ corresponds to the
Horodecki diagram. The generalization here is that $\alpha$, and
thus the entanglement, is arbitrary.

When we take the partial transpose we will get a new tetrahedron,
consisting of the states $\rho^{T_b}=\sum_i w_i{\rho_i}^{T_b}$.
The partial transposition is actually given by a reflection in an
11-dimensional plane in the 15-dimensional space. The new
tetrahedron will not, except for $\alpha=0$ and $\alpha=\pi/4$,
lie in the same 3-dimensional subspace as the tetrahedron with the
basis states at the vertices. But we will see that there is a
plane of fixed points; the two tetrahedra intersect in a
2-dimensional plane. The non-separable states will, through the
reflection, pass to unphysical regions, i.e. have negative
eigenvalues, while the reflection of separable states will give
separable states, with non-negative eigenvalues. At the boundary
between these regions the determinant of the partially transposed
matrix will vanish:
\begin{equation}
  \mathrm{det} \: \rho^{T_b} =0 \ .
\end{equation}
We will use this to find the boundary between separable and
non-separable states in the original tetrahedron. Through a
straightforward calculation equation~(4) is seen to give two
possibilities:
\begin{eqnarray}
  \cos^2\!\alpha \: \sin^2\!\alpha \left( \left( w_1-w_2 \right)^2-\left( w_3-w_4 \right)^2 \right)+ w_1w_2 \!\!\!&=&\!\!\! 0 \ , \\
  \cos^2\!\alpha \: \sin^2\!\alpha \left( \left( w_1-w_2 \right)^2-\left( w_3-w_4 \right)^2 \right)- w_3w_4 \!\!\!&=&\!\!\! 0 \ .
\end{eqnarray}

Now we introduce the cartesian coordinates (see figure 1a)
\begin{equation}
  x = \frac{1}{2}(w_1-w_2) \ , \quad
  y = \frac{1}{2}(w_3-w_4) \ , \quad
  z = \frac{1}{2\sqrt{2}}(w_3+w_4-w_1-w_2) \ .
\end{equation}
Equations (5) and (6) give the surfaces
\begin{eqnarray}
  \left(z+\frac{1}{2\sqrt2} \right)^2 \!\!\!\!\!&=&\!\!\!
  8\cos^2\!\alpha \: \sin^2\!\alpha\ y^2 + 2\left( 1-4\cos^2\!\alpha \: \sin^2\!\alpha \right)x^2 \ , \\
  \left(z-\frac{1}{2\sqrt2} \right)^2 \!\!\!\!\!&=&\!\!\!
  8\cos^2\!\alpha \: \sin^2\!\alpha\ x^2 + 2\left( 1-4\cos^2\!\alpha \: \sin^2\!\alpha \right)y^2 \ .
\end{eqnarray}

For $0<\alpha<\pi/4$ we have intermediate entangled basis states.
Then (8) and (9) are the equations of two elliptical cones with
apices in $(0,0,\mp 1/2\sqrt2)$ and their axes along the z-axis.
The cones cut the $xy$-planes in ellipses. For one of the cones
the major axis is along the x-axis and for the other along the
y-axis (which one depends on $\alpha$). In the plane $z=0$ the
cones always intersect at the edges of the tetrahedron. These
cones together with the sides of the tetrahedron confine the
region of all separable states, as shown in figure~1. This region
decreases when $\alpha$ increases, i.e. when the entanglement of
the basis states increases. For $\alpha=\pi/8$ we have the special
case with circular cones.

\begin{figure}
  \begin{center}
  \includegraphics{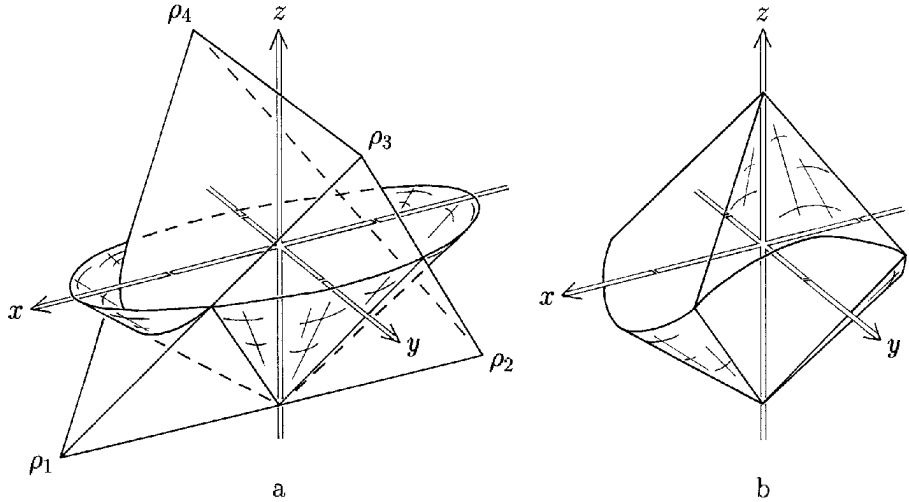}
  \end{center}
  \caption{\small a) The tetrahedron with the basis states at the
  vertices and one of the elliptic cones dividing separable states
  from non-separable states. b) The region of separable states.}
\end{figure}

When $\alpha=0$ or $\alpha=\pi/4$ the two cones will degenerate to
four planes; the major axes of the ellipses goes to infinity while
the minor axes stays finite. For $\alpha=0$ these planes coinside
with the four sides of the tetrahedron. Thus every mixture is
separable. This is also obvious because the vertices are the
product states, which gives $\rho=\sum_i w_i \rho_i$ on the form
(1). We have $\rho^{T_b}=\rho$\,; the tetrahedron lies in the
plane of reflection.

When $\alpha=\pi/4$ we have the Bell basis at the vertices and
this tetrahedron corresponds to the Horodecki diagram (but the
distances differ with a factor $2\sqrt2$ [2,3]). The four planes
given by equations (8) and (9) are the sides of the inverted
tetrahedron and we have obtained the ``stella octangula'', with
the separable states in the octahedron.

Among the separable states some are fixed points with respect to
Peres' transformation, i.e. they fullfill
\begin{equation}
  \rho^{T_b}=\rho \ .
\end{equation}
As already noted every point is a fixed point when $\alpha=0$. But
for $0<\alpha\leq\pi/4$ (10) gives $w_1+w_4=w_2+w_3=1/2$\,, or in
our cartesian coordinates: $x=y$\,. This plane are the fixed
points. In the tetrahedron this is a square. For the Bell basis
case taking the partial transpose is a reflection in this plane.

We have met with two properties of the boundary of density
matrices mentioned by Harriman~[5]: At every boundary point there
are some directions in which the boundary is curved, but in other
directions it is straight (except for pure states). This is
clearly seen in the cases with elliptical cones.

The possibility to illustrate a 15-dimensional space with
3-dimensional intersections is of course limited, but these pictures
give at least some understanding of how things work. When the
entangelment of the basis states at the vertices of the tetrahedron
varies we sweep over different intersections and we see how the region
of separable states decreases when the entanglement of the basis
states increases. For another attempt to illustrate how the separable and
non-separable states are located in the set of all density
matrices, see [6].

I am grateful to Ingemar Bengtsson for suggesting the problem
studied in this paper, for many helpful discussions and for his
patience with me.

\vspace{10mm}
\noindent {\large \textbf{References}} \\[4 pt]
{\small
  $[1]$\ \ A. Peres, Phys. Rev. Lett. \textbf{77} (1996) 1413.\\
  $[2]$\ \ R. Horodecki and M. Horodecki, Phys. Rev. A \textbf{54} (1996) 1838.\\
  $[3]$\ \ P. K. Aravind, Phys. Lett. A \textbf{233} (1997) 7.\\
  $[4]$\ \ M. Horodecki, P. Horodecki and R. Horodecki, Phys. Lett A 
  \textbf{233} (1996) 1.\\
  $[5]$\ \ J.E. Harriman, Phys. Rev. A \textbf{17} (1978) 1249.\\
  $[6]$\ \ F. Verstraete, J. Dehaene and B. De Moor, e-print archive: quant-ph/0107155.\\
}

\end{document}